\documentclass[letterpaper,twocolumn,10pt]{article}
\usepackage{usenix2019_v3}

\usepackage{tikz}
\usepackage{amsmath}

\usepackage{filecontents}

\usepackage{cite}
\usepackage{amsmath,amssymb,amsfonts}
\usepackage{algorithmic}
\usepackage{graphicx}
\usepackage{textcomp}
\usepackage{xcolor}
\usepackage{booktabs} 
\usepackage{graphicx}
\usepackage{color}
\usepackage{xspace}
\usepackage{multirow}
\usepackage{amssymb}
\usepackage{amsmath}
\usepackage{tabularx}
\usepackage{ltxtable}
\usepackage{tablefootnote}
\usepackage{fancyhdr}
\usepackage{etoolbox}
\usepackage[title]{appendix}
\usepackage{comment}

\usepackage{array}
\newcolumntype{P}[1]{>{\centering\arraybackslash}p{#1}}
\newcolumntype{M}[1]{>{\centering\arraybackslash}m{#1}}

\newcommand{\cut}[1]{}

\newcommand{\sysname}{{\sc Ropnn}\xspace}

\begin{document}

\date{}
\title{
\bf{\sysname: Using Deep Neural Networks to Accurately \\ Detect ROP Payloads}  
}

\author{
Xusheng Li$^1$
\and Zhisheng Hu$^1$
\and Haizhou Wang$^1$
\and Yiwei Fu$^1$
\and Ping Chen$^2$
\and Minghui Zhu$^1$
\and Peng Liu$^1$
}

\date{%
    $^1$The Pennsylvania State University\\%
    $^2$JD.com American Technologies Corporation\\[2ex]%
}

\maketitle
\setcounter{page}{1}

\pagestyle{plain}


\vspace{-10mm}
\begin{abstract}


Return-oriented programming (ROP) is a code reuse attack that chains short snippets of existing code to perform arbitrary operations on target machines. Existing detection methods against ROP exhibit unsatisfactory detection accuracy and/or have high runtime overhead. 

In this paper, we present \sysname, which innovatively combines address space layout guided disassembly and deep neural networks to detect ROP payloads. The disassembler treats application input data as code pointers and aims to find any potential gadget chains, which are then classified by a deep neural network as benign or malicious. Our experiments show that \sysname has high detection rate (99.3\%) and a very low false positive rate (0.01\%). \sysname successfully detects all of the 100 real-world ROP exploits that are collected in-the-wild, created manually or created by ROP exploit generation tools. Additionally, \sysname detects all 10 ROP exploits that can bypass Bin-CFI. \sysname is non-intrusive and does not incur any runtime overhead to the protected program. 
\end{abstract}


\section{Introduction}\label{sec:intro}

Due to broad deployment of W$\oplus$X or Data Execution Prevention (DEP)~\cite{DEP,ExecShield}, code injection attacks (e.g., shellcode injection) are no longer very viable. Code reuse attacks, especially return-oriented programming (ROP) attacks~\cite{JOP2,JOP1,Q-ROP,ROP1}, have recently risen to play the role code injection attacks used to play. A ROP attack proceeds with two phases. First, the attacker identifies a set of particular machine instruction sequences (i.e., ``gadgets") that are elaborately selected from a binary executable or a shared library. Each gadget typically ends in a return instruction.  
Second, enabled by exploiting a (buffer overflow) vulnerability, the attacker overwrites part of the stack with the addresses of these gadgets; the addresses and register operands are placed at  particular locations (on the stack) so that these gadgets will be executed sequentially if the control flow is directed to the first one. By chaining gadgets together, the attacker is often able to perform arbitrary operations on the target machine \cite{ROP1}. 

Since ROP attacks are a major threat to business-critical server programs, extensive researches have been conducted to defend against ROP attacks. The existing defenses focus on two perspectives, namely prevention and detection. ROP prevention methods aim to make launching a ROP attack itself very difficult. For example, 
ASLR (Address Space Layout Randomization) randomizes the memory addresses to make it difficult for an attacker to accurately locate any ROP gadgets. However, ASLR suffers from various forms of information leakage~\cite{strackx2009breaking, JITROP, seibert2014information}, code inference~\cite{snow2016return} and indirect profiling attacks~\cite{rudd2017address}, which significantly undermine the protection. 

On the other hand, ROP detection methods aim to catch the ROP attack in action and then stop it. For example, CFI (Control Flow Integrity) checks whether the control flow transfer violates the CFG (Control Flow Graph). If so, the program is terminated. Existing ROP detection methods can be divided into six classes: (1A) Heuristics-based detection~\cite{DROP,ropecker,kbouncer}. (1B) Fine-grained CFI~\cite{akritidis2008preventing,payerfine,CCFI}. (1C) Signature-based detection~\cite{tanaka2014n}. (1D) Speculative code execution~\cite{polychronakis2011rop}. (1E) Searching code pointers in the data region~\cite{stancill2013check}. (1F) Statistical-based detection~\cite{pfaff2015learning,elsabagh2017detecting}.


Unfortunately, these methods are still quite limited in meeting four highly-desired requirements:
\textbf{(R1)} high detection rate for ROP attacks;
\textbf{(R2)} close to zero false positive rate; \textbf{(R3)} acceptable runtime overhead; \textbf{(R4)} minimal changes to the protected programs and the running environment. 

In fact, (a) Class 1A and 1C detection methods could result in low detection rates; specific heuristics or signatures are found not very hard to bypass.   (b) Class 1B and 1D detection methods could cause substantial runtime overhead. (c) Class 1A, 1B, 1D, and 1F detection methods may cause substantial changes to existing (legacy) application software and even the running environment, thus they are not transparent. (Please refer to Section~\ref{sec:existing} for detailed discussion.) As a result, none of the existing ROP detection methods satisfy the above four requirements. 

\begin{table}[tbp]
  \centering
  \scriptsize
  \begin{tabular}{|l|l|l|l|l|}
  \hline
  Detection Methods & R1 & R2 & R3 & R4  \\ \hline
  Heuristic-based  & $\times$ & \checkmark  & \checkmark & $\times$ \\ \hline
  Control Flow Integrity (CFI)  & \checkmark & \checkmark & $\times$ & $\times$ \\ \hline
  Signature-based  & $\times$ & \checkmark & \checkmark & \checkmark \\ \hline
  Speculative code execution & \checkmark  & \checkmark & $\times$ & $\times$  \\ \hline
  Searching code pointers in the data region & $\times$ & \checkmark & \checkmark & \checkmark \\ \hline
  Statistical-based detection & $\times$ & \checkmark & \checkmark & $\times$ \\ \hline
  
  \end{tabular}
 \caption{Limitations of existing detection methods against ROP attacks.}
\label{tab:detecion}
\vspace*{-3mm} 
\end{table}


In recent years, deep neural network sees applications in the security field, e.g., fuzzing~\cite{bottinger2018deep}, log analysis~\cite{du2017deeplog}, memory forensic~\cite{song2018deepmem}, 
etc. Deep neural network has several clear advantages over traditional machine learning methods, for example it provides better accuracy than conventional models like Support Vector Machine (SVM); it does not require expert knowledge to set thresholds for classification (detection) criteria; it does not require traditional feature engineering and can be trained end-to-end using minimally preprocessed data. 

In this paper, we study whether these advantages could be leveraged to address the aforementioned limitations of existing ROP payload detection methods. In particular, we propose a new ROP payload detection method, \sysname, which is the first to satisfy all of the above four requirements via deep learning. 
Regarding how \sysname works, firstly, our method is used as a ``classification engine'' to  build a new kind of network Intrusion Detection System (IDS). It can be deployed in the same way as a conventional network IDS such as Snort. Secondly, once deployed, our method works in the following manner: when network packets arrive and a reassembled protocol data unit (PDU) is obtained, our method takes two steps. (Step 1) Our method does ASL-guided PDU (i.e., application input data) disassembly and obtains a set of potential gadget chains. 
(Step 2) The potential gadget chains obtained in Step 1 are fed into a neural network classifier. The classifier identifies each potential gadget chain as either ``ROP payload" or ``benign data''. 

As we will show shortly in Section~\ref{sec:evalution}, \sysname achieves very high detection rate (99.3\%) and very low false positive rate (0.01\%), which satisfy R1 and R2. Meanwhile, since \sysname can be deployed on a separate machine other than the protected server, it requires no changes to the protected program or the running environment (R4). ROPNN also has no runtime overhead for the protected program (R3), which is an advantage over many other methods. 

Despite the successful applications of deep neural network in other security problems~\cite{bottinger2018deep, du2017deeplog, song2018deepmem}, \sysname still faces several unique domain-specific challenges. Firstly, a deep neural network must be trained with proper data. Since ROP payloads only contain addresses of ROP gadget chains (please refer to Section~\ref{sec:rop}), we should not train a classifier to directly distinguish ROP payloads from benign data. Otherwise, the signal-to-noise ratio is so low that the accuracy can be very poor. Instead, we propose ASL-guided disassembly (Section~\ref{sec:asl}) and create gadget-chain-like instruction sequences based on the addresses identified in benign data. In section~\ref{sec:real}, we also propose a viable method to generate sufficient real gadget chains. Simply put, the two datasets are both a set of instruction sequences and we train a classifier to distinguish the two. Also, to obtain comprehensive and representative benign training samples, we do ASL-guided disassembly on TB-level amount of raw input data (HTTP traffic, images, PDFs). We obtain 26k-105k benign training samples for different programs. 

Secondly, we need to design a customized deep neural network for the detection of ROP payloads. We propose to use a convolutional neural network (CNN) as our classifier as it is good at capturing spatial structure and local dependencies. This corresponds to the nature of a ROP gadget chain that gadgets are chained with orders and adjacent instructions in a chain have meaningful connections with each other. These orders and connections in turn indicate whether the instruction sequence is indeed a real gadget chain or is formed merely due to the coincidental addresses in the data. 

In summary, the contributions of this paper are as follows: 


\begin{itemize}
\item We propose \sysname, a novel method for a network IDS to use in detecting ROP payloads. It combines ASL-guided disassembly and deep learning to classify reassembled PDU into either ``ROP payload" or ``benign data". 

\item To the best of our knowledge, \sysname is the {\em first} intrusion detection method that applies deep learning to mitigate the threat of ROP attacks. It sheds light on the applicability of deep learning to software system security problems.  

\item We have designed and implemented \sysname on Linux. We test it with several programs (e.g., Nginx). The evaluation results show that \sysname achieves very high detection rate (99.3\%) and very low false positive rate (0.01\%). More importantly, it can detect real-world ROP exploits collected in-the-wild and generated by widely-used exploit generation tools. We collect and create 100 real-world ROP exploits for a total of five vulnerable programs and \sysname successfully detects all of them. Additionally, \sysname detects all 10 ROP exploits that can bypass Bin-CFI. Meanwhile, \sysname does not require changes to the program or the running environment. It also incurs no runtime overhead to the protected program. 


\end{itemize}


\section{Background}
\subsection{ROP Attacks}\label{sec:rop}

\begin{figure}[t]
\centering
  \includegraphics[width=1\columnwidth]{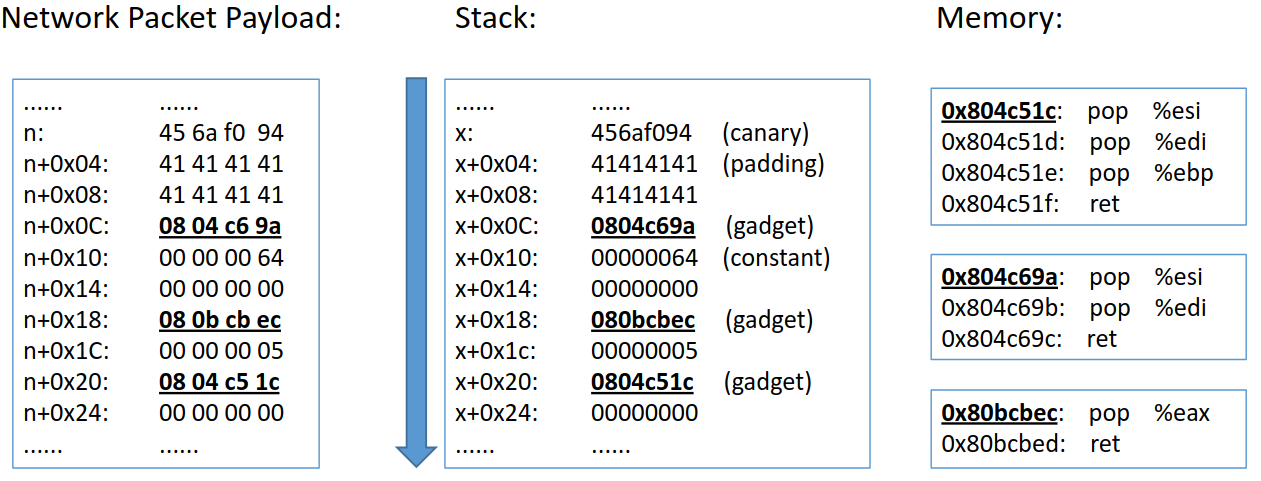}
  \caption{Relationship between network packet payload, stack layout, and address space layout}~\label{fig:brop}
\vspace*{-5mm} 
\end{figure}

Despite years of effort to mitigate the threat, ROP attack remains a widely used method to launch exploits~\cite{attack1, attack2, attack3, attack4}. In a typical ROP attack, an attacker carefully crafts a gadget chain and embeds the addresses of the gadgets in a network packet. This packet exploits a vulnerability in the server program and causes the gadget to be executed one by one. Figure \ref{fig:brop} shows the relationship of the network packet payload, the overflowed stack, and the program's memory address space layout. In fact, this comes from the blind ROP attack \cite{ROPonNginx}, which exploits a stack buffer overflow vulnerability in Nginx 1.4.0. As we can see, the addresses of gadgets (e.g., 0x804c51c, 0x804c69a) are embedded in the network packet. This relationship is an intrinsic characteristic of ROP attacks and it motivates us to search for addresses of potential gadgets in input data and then disassemble the corresponding instructions. 

On the other hand, since only the address of ROP gadgets are contained in the ROP payloads, we should  not directly label ROP payloads and benign data and then train a classifier on the two -- because the signal-to-noise ratio is too low and it is difficult to train an accurate classifier. 

\subsection{Existing Detection Methods against ROP Attacks}\label{sec:existing}


In this subsection, we classify the existing detection methods against ROP attacks into six categories and explain why they fail to meet the four requirements set in Section~\ref{sec:intro}. 

Heuristic-based detection methods look for abnormal execution patterns.
DROP~\cite{DROP} checks whether the frequency of executed return instructions exceeds a certain threshold. kBouncer~\cite{kbouncer} and ROPecker~\cite{ropecker} detect abnormal pattern of indirect branches.
Although these methods can detect conventional ROP attacks, they leverage certain heuristic-based thresholds as critical detection criteria. G{\"o}kta{\c s} et al. point out in~\cite{JX4} that they can be bypassed by carefully constructed gadget chains which violate the defender's assumptions. 

Control Flow Integrity (CFI) ~\cite{CFI00,CFI9,CFI4\cut{,CFI5},CFI3,CFI6} involves two steps, i.e., detection and response. CFI detects whether any control flow transfer violates the pre-defined control flow graph (CFG). If so, it terminates the program. Theoretically, CFI can detect all ROP attacks since they inevitably violate the CFG. However, due to the difficulty of point-to analysis, obtaining a perfect CFG is difficult (if not impossible)~\cite{burow2017control}. In practice, coarse-grained CFI implementations adopt an approximated CFG, which is not as secure as it should be. Consequently, attackers can leverage the extra edges in the CFG to bypass them~\cite{ROP11,CFI4,JX2,JX5}. On the other hand, fine-grained CFI provides a strong security guarantee at the cost of significantly high runtime overhead; e.g., 19\% in~\cite{payerfine} and 52\% in~\cite{CCFI}, which is impractical. Meanwhile, CFI requires either instrumentation to the program binary or modifications to the compiler.

There are also signature-based detection methods against ROP, i.e., n-ROPdetector \cite{tanaka2014n}. n-ROPdetector checks whether a set of addresses of API functions appear in network traffic. While this method can detect some ROP attacks, 
the address of APIs can be masqueraded to evade the detection. For example, an attacker can represent an address as the sum of two integers, and calculate the address at runtime.

Speculative-code-execution-based detection (ROPscan~\cite{polychronakis2011rop}) searches network traffic or files for any code pointers 
and starts speculative code execution from the corresponding instruction. If four or more consecutive gadget chains can be successfully executed, ROPscan considers the input data as a ROP exploit. 
Note the threshold four is empirically determined in the experiment; however, check my profile \cite{stancill2013check} finds that benign data can produce a gadget chain that has up to six gadgets. Meanwhile, real gadget chains need not to be very long to be useful. Thus the selection of the threshold is challenging. In fact, this also motivates the use of deep neural network as a classifier because no such thresholds need to be provided. 



Check my profile \cite{stancill2013check} statically analyzes the data region and looks for any code pointers that form potential gadget chains. 
To avoid false positives, it checks whether the gadget chain eventually calls an API function -- which most useful exploits will do. Unfortunately, due to the nature of static analysis, the detection can be bypassed when the address of API function is masqueraded, e.g., as a sum of two integers. 


Statistical-based detection methods \cite{elsabagh2017detecting,pfaff2015learning} 
first extract certain meta-information (e.g., micro-architectural events) about the program execution and then train a statistical machine learning model (e.g., SVM) to distinguish ordinary execution from ROP gadget chain execution. 
They can detect various types of ROP attacks. However, they require instrumentation to acquire such information. Moreover, they use relatively small training dataset, which leaves plenty of spaces for improvement. For example, EigenROP \cite{elsabagh2017detecting} achieves 80\% detection rate and 0.8\% false positive rate, whereas we get 99.3\% detection rate and 0.01\% false positive rate. Note our false positive rate means 0.01\% of the potential gadget chains, and the number of such chains is significantly smaller the number of program inputs (see Section~\ref{sec:evalution}), so our accuracy is much better than EigenROP.

\section{Overview}\label{sec:overview}
\begin{figure}[t]
\centering
\includegraphics[width=1\linewidth]{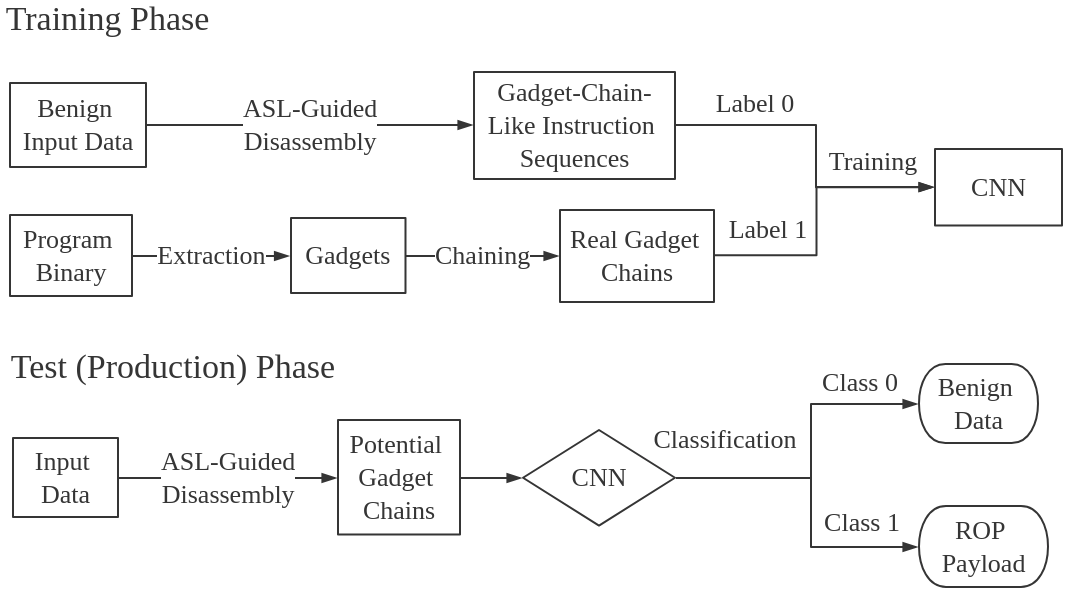}
 \vspace*{-6mm}
    \caption{Workflow of \sysname}
    \label{fig_ropnn}
    \vspace*{-3mm}
\end{figure}

If a PDU does not contain any 4-byte (8-byte) value which points to an instruction in a 32-bit (64-bit) system, the PDU is certainly not a ROP payload. Otherwise, the PDU is suspicious. A main challenge for detecting ROP payloads is that most suspicious PDUs are actually benign. Now we need to accurately judge if a suspicious PDU is a ROP payload or not. Since creative zero-day ROP attack scripts could be invented at any time, no IDS knows all the essential features of ROP attacks.  This key observation reveals a fundamental limitation of the ROP detection methods that rely on known features (e.g., abnormal patterns of indirect branches, addresses of API functions being contained in a PDU); it also indicates that an evasion-resilient intrusion detection method should be {\em feature-free}. A main reason on why we leverage deep learning is that such learning models do not require any attack features to be explicitly identified. 

Figure \ref{fig_ropnn} illustrates the workflow of our proposed \sysname. Overall, our approach has two phases: the training phase and the production phase. During the training phase, we first collect two training datasets, namely the gadget-chain-like instruction sequences and the real gadget chains. We get the gadget-chain-like instruction sequences by Address Space Layout (ASL)-guided disassembly (of the protected program's memory dump) based on the valid addresses identified in the benign input data. We extract real gadget chains from the protected program by chaining individual gadgets. Note by instruction we mean both opcode and operands (if any). 
We then 
train a deep neural network to classify the two datasets. Specifically, we choose to use a three-layer convolutional neural network (CNN) and represent the instructions byte-by-byte using one-hot encoding (Section \ref{sec:one-hot}).

During the production phase, after a PDU is obtained (re-assembled), we first identify any valid addresses contained in the PDU. Then we use the addresses to perform ASL-guided disassembly, which will result in a set of potential gadget chains. After that, we send each of them to the trained CNN. A warning is raised if any of these potential gadget chains is classified as real ROP attack. Conversely, if the ASL-guided disassembly does not produce a potential gadget chain, or the chain is classified as benign by the CNN, the input data is considered benign.

Note we do need a classifier as part of the \sysname because if we treat all of the inputs that have addresses in it as ROP payloads, we can cause high false positive rate and substantial denial of service. Meanwhile, the ASL-guided disassembly is also necessary because otherwise the signal-to-noise ratio is very low. An HTTP request can be as large as a 2MB image. However, suppose a ROP gadget chain has 20 gadgets, then only 20 addresses are present in the payload, which means 80 bytes. Viewing ROP payloads as a ``haystack'', the addresses embedded in by an attacker are in fact some ``needles''. If we treat a whole payload as a training sample, then any trained neural network will very likely be primarily capturing the features of the haystack instead of the needles.  If this is the case, a very low classification accuracy is unavoidable. 

We choose CNN as the classifier because ASL-guided disassembly outputs instruction sequences which have strong spatial structures and local dependencies. The appearance of pairs of instructions and the orders of instructions could indicate different nature of an instruction sequence. CNN can represent the abstract nature (e.g., benign or real gadget chain) into compositions of simpler concepts (e.g., locations of instructions), which is important for the classification. 



This unique combination of ASL-guided assembly and CNN enables us to solve the ROP payload detection problem without any software instrumentation. The only thing that the \sysname needs to know about is a memory dump of the protected program. If we train a classifier to distinguish ordinary execution trace from ROP gadget chain execution trace, we need software instrumentation to monitor the execution trace, which is against our goal.

\section{ASL-Guided Disassembly and Gadget-Chain-Like Instruction Sequences Generation}\label{sec:asl}

This section describes the details about the ASL-guided disassembly. ASL-guided  disassembly treats bytes in data as addresses and checks if they point to gadget-like or gadget-chain-like instruction sequences. In the training phase, we use ASL-guided disassembly to collect gadget-chain-like instruction sequences as training data for the neural network. In the production phase, we use ASL-guided disassembly to identify any potential gadget chain in input data. \cut{We give an example of the ASL-guided disassembly at the end of this section.}  




\subsection{Disassembly of Individual Addresses}\label{sec:disassem}

We first create a memory dump of the protected program. The memory dump contains the addresses and contents of all memory pages that are marked as executable. Oftentimes, these pages are mapped from the {\tt text} segment of the protected program or any loaded shared libraries (e.g., libc.so, etc). The ROP gadgets must fall inside the memory dump, otherwise it is not executable and the attack will fail. 


Then we consider every four bytes (on 32-bit system)\footnote{In this paper, we only consider 32 bit systems and leave it for a future work to experiment with 64 bit systems.} in the input data as an address. If an address does not appear inside any one of the dumped pages, we ignore it because it cannot be the start of a ROP gadget. Here we do not limit our search space to non-randomized modules because attackers can bypass the ASLR in multiple ways. 

We start disassembling from any identified addresses using Capstone~\cite{capstone}. The disassembling can stop in two ways: (1). An invalid or privileged instruction is encountered or the disassembly reaches the end of code segment. In this case, the current address is ignored in subsequent analysis. (2). An indirect branch, i.e., {\tt ret}, {\tt jmp}, or {\tt call} is encountered. Then the instruction sequence (from the instruction at the starting address all the way to the indirect branch) is considered a gadget-like instruction sequence. 

\subsection{``Chaining'' of Gadget-Like Instruction Sequences}\label{sec:chain}

Having obtained the set of gadget-like instruction sequences, we need to figure out how they could be chained together, in a similar way as an attacker chains gadgets into a gadget chain. Specifically, if we find an address at offset $n$ in the data which points to a gadget-like instruction sequence, we check if any one of the next ten addresses, i.e., address at $n+4$, $n+8$, $n+12$,...,$n+40$ in the data also points to a gadget-like instruction sequence. If so, we ``chain'' it together with the previous gadget-like instruction sequence and repeat the process. Otherwise, we end the ``chaining'' process. 

For any ``chain'' that has at least two addresses, we collect the corresponding gadget-like instruction sequences and concatenate them into a gadget-chain-like instruction sequence. Note the maximum ten is determined according to the observation that most gadgets in our dataset only {\tt pop} less than five integers from the stack to the registers, and all of them pop less than ten integers; so ten is sufficient to capture the next address in a ``chain''. In other words, a ROP attack cannot spread the addresses of gadget chains in the payload arbitrarily; otherwise, even if the control flow is successfully hijacked, the subsequent gadgets will not execute one by one -- because upon return from the previous gadget, the address of the next address is not on the top of the stack. 

When we look for the next ``chain'', we skip the addresses and the corresponding instruction sequences that are already part of a ``gadget chain''. For example, if a ``chain'' contains five ``gadgets'', next we start from the sixth address and repeat the ``chaining'' process. We repeat the process on every address to collect all possible gadget-chain-like instruction sequence. In this way, we get the first training dataset. 

To efficiently implement the above algorithm, we start multiple parallel threads to analyze different addresses. Moreover, if an address is already examined and found to be pointing to a gadget-like instruction sequence, we cache it in a global table. In this way, the disassembly and analysis is not repeated on the same address. Besides, we also process multiple inputs simultaneously to utilize all CPU cores. 

The reader may take Figure~\ref{fig:brop} as an example of the ASL-guided disassembly during the production phase. Suppose we start from offset n of the data, we first encounter a candidate address 0x456af094. We find this address is not inside the memory dump, which indicates it is not mapped or not marked as executable in the protected program, so we move on to the next 4 bytes starting at offset n+0x04. The next two addresses are 0x41414141 and they both do not lead to a potential gadget chain. Next, we move to offset n+0x0c and process 0x0804c69a. Note this address is inside the memory dump and we start disassembling from it. 
The result is a potential gadget with three instructions: \texttt{pop esi; pop edi; ret;}. We continue the process and then identify two other addresses (0x080bcbec, 0x0804c51c) and their corresponding instructions. Eventually, we end up with a potential gadget chain with three gadgets (\texttt{pop esi; pop edi; ret; pop eax; ret; pop esi; pop edi; pop ebp; ret;}). In the training phase, however, when we collect gadget-chain-like-instruction sequences, we process benign input data using the same approach. 

It is noteworthy that we start ASL-guided disassembly from EVERY byte of the input data. This is because we are dealing with data, so code or memory alignment actually does not apply, and any four-bytes can be an address. In fact, in Figure~\ref{fig:brop}, we also treat the four-byte data at n+1 (0x6af09441), n+2(0xf0944141) and n+3(0x94414141) as addresses and start ASL-guided disassembly. Though they do not lead to gadget-like instruction sequences and are ignored in the subsequent analysis.

\section{Real Gadget Chain Generation}\label{sec:real}

The real gadget chain dataset is created by chaining real individual gadgets together. 
There are several existing tools to automate gadget chains generation; e.g., rp++~\cite{rpplus2012}, ROPgadget~\cite{ROPgadget2015}, ropper~\cite{ropper2016}, PSHAPE~\cite{Follner2016}, ROPER~\cite{roper2017}. However, the existing tools cannot be directly used to generate the real gadget chain dataset due to three main reasons: 1. The number of generated gadget chains is small. Many existing tools usually build ROP exploits for one specific scenario;  e.g., {\tt execve} or {\tt mprotect}, which leads to a small number of real gadget chains. 2. Existing tools usually use gadgets whose lengths are as short as possible to reduce side-effects on other registers, the stack, or flags. This makes the dataset not comprehensive. 3. The generated gadget chains might cause crashes due to accesses of unmapped memory. 


We decide to generate the real gadget chains in a new way. Our idea is to first generate a lot of candidates, and then filter out those invalid ones.  
We use ROPgadget~\cite{ROPgadget2015} to extract individual gadgets from the program binary. Then the gadgets are added to the chain in such a way that every register is initialized before the chain dereferences it, as the execution may otherwise lead to a crash. To avoid crashes, we have to solve the side effect of gadgets. Take two gadgets ``mov [esi], 0x1; ret;'' and ``mov eax, 0x1; jmp [esi];'' for instance. There exists a side effect caused by the gadget ``mov eax, 0x1; jmp [esi];'' unintendedly changing the value of EAX, and a register usage conflict for ESI, which is used for setting a memory as 0x1 and setting the target address for jump instructions. To solve the side effects, we remove all the gadgets that contain the memory usages, and make sure that no two gadgets read one register without write operation between them. To make the dataset more comprehensive, we also combine both short and long gadgets.

After that, we use the CPU emulator Unicorn~\cite{unicorn} to validate the generated gadget chains. We first analyze how the individual gadgets interact with the stack pointer. For example, if a gadget pops two integers into registers, we know that after the execution of this gadget, the new {\tt esp} value will become {\tt esp+0xc}. We arrange the addresses of gadgets on the stack inside Unicorn according to their stack interaction. We then start emulation from the first gadget and observe if the gadgets can be executed one-by-one correctly. If not, this gadget chain is filtered out. One exception to the emulation is all function calls (e.g. {\tt call eax}) are assumed successful and the corresponding function call is skipped. 
In this way, we make sure the generated gadget chains are all valid (they are not necessarily useful for attackers).

We can generate a huge amount of real gadget chains in this way. However, if we contain too many real gadget chains in the training data, the neural network will tend to classify more samples as a real gadget chain, which can lead to higher false positives. To avoid this, we generate the same amount of real gadget chains as gadget-chain-like instruction sequences. 

Furthermore, we make sure the length (in bytes) distribution of the real gadget chain datasets is similar to that of the gadget-chain-like instruction sequence dataset; otherwise, even if the classifier does a good job to distinguish the two datasets, it may leverage the length information too much rather than learn anything about the data.


\section{Neural Network Classification}\label{sec:nn}
We formulate the ROP payloads detection problem as a classification problem where our goal is to discriminate ROP gadget chains from gadget-chain-like instruction sequences.
\cut{However, deep neural networks face }We now need to tackle two main challenges in order to successfully apply deep neural networks. First, deep neural networks require a huge amount of training data to perform well. For example, deep neural networks perform worse than some traditional classifiers when the dataset has less than 100 samples \cite{DNUDPSD}. As mentioned in Section~\ref{sec:chain} and Section~\ref{sec:real}, it is challenging to get a large number of gadget-chain-like instruction sequences. In fact, we use ASL-guided disassembly to disassemble TB level of raw input data to generate enough instruction sequences (please refer to Section~\ref{sec: big_data}). Second, different types of deep neural networks are suitable for different types of data; e.g., convolutional neural networks (CNN) work well in image classification since the data has spatial structures~\cite{ciresan2011flexible}, and recurrent neural networks (RNN) with long short-term memory (LSTM)~\cite{hochreiter1997long} can deal with temporal structures in spoken language. It is very important to design the correct architecture of the deep neural network according to the nature of instruction sequence data in our task. 

\subsection{Data Representation and Preprocessing}\label{sec:one-hot}

In order to make our two datasets (i.e., the gadget-chain-like instruction sequences and the real gadget chains) tractable for the neural network, we need to convert every instruction in them into numerical data. Recall that instructions are binary data by nature. We first convert every byte in an instruction sequence into its hex value (0-255). For example, the instruction sequence ``mov eax, 0x1; pop edx; ret'' is converted to ``[0xb8, 0x01, 0x00, 0x00, 0x00, 0x5a, 0xc3]''. 

We can simply scale these values to 0-1 by a division of 255 and then input them into the neural network. However, this is inappropriate for our data. Since neural networks multiply the input data with certain weight parameters, such representation leads to an implicit relative order between different byte values. For example, in the above example, the neural network implicitly assumes the ``ret'' (0xc3) is larger than ``pop edx'' (0x5a), which is meaningless regarding instructions. 

To address this problem, we instead use one-hot encoding. We represent each byte value by a 256 $\times$ 1 vector, with all but one positions to be zero, and the position that corresponds to the byte value to be one. For example, the ``ret'' (0xc3, 195) will be represented by \{0,...,0,1,0,...,0\}, where we have 195 zeros in front of the one, and 60 zeros behind the one. 

In other words, an instruction sequence that has $n$ bytes is represented by $X = \{\overrightarrow{X_1}, \overrightarrow{X_2}, ..., \overrightarrow{X_n}\}$, where $\overrightarrow{X_i}$ is a $256\times1$ one-hot vector. Alternatively, one can view the input as a $n$ by 256 matrix. 

Since instruction sequences usually have different lengths, padding is applied to make them have the same length. We first find the longest instruction sequence (in bytes). For shorter ones, we append the one-byte {\tt nop} (0x90) instruction at the end of them until they all reach the same length. 

An alternative way to preprocess the data is to do word embedding~\cite{lai2016generate}, which is widely used in Natural Language Processing (NLP). Word embedding maps every word in the data to a fixed length vector. 
In the evaluation section, we show that embedding, compared to one-hot encoding, provides similar accuracy while requires longer training time. Consequently, one-hot encoding is more suitable for our task.

\subsection{Architecture of the Deep Neural Network}

\begin{figure}[t]
\centering
\includegraphics[width=1\linewidth]{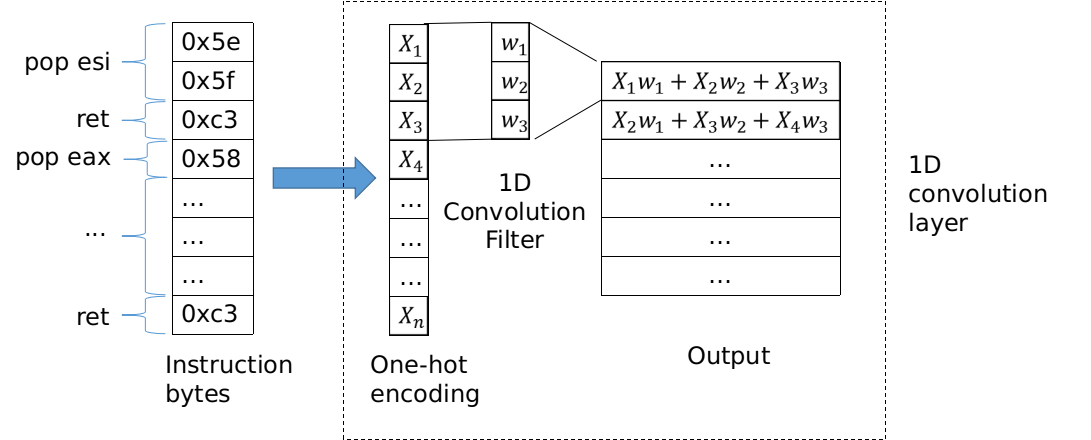}
    \caption{Data Representation and the first 1D Convolution Layer}
    \label{fig:conv}
    \vspace*{-5mm}
\end{figure}

\begin{figure*}[h]
\centering
\includegraphics[width=0.7\linewidth]{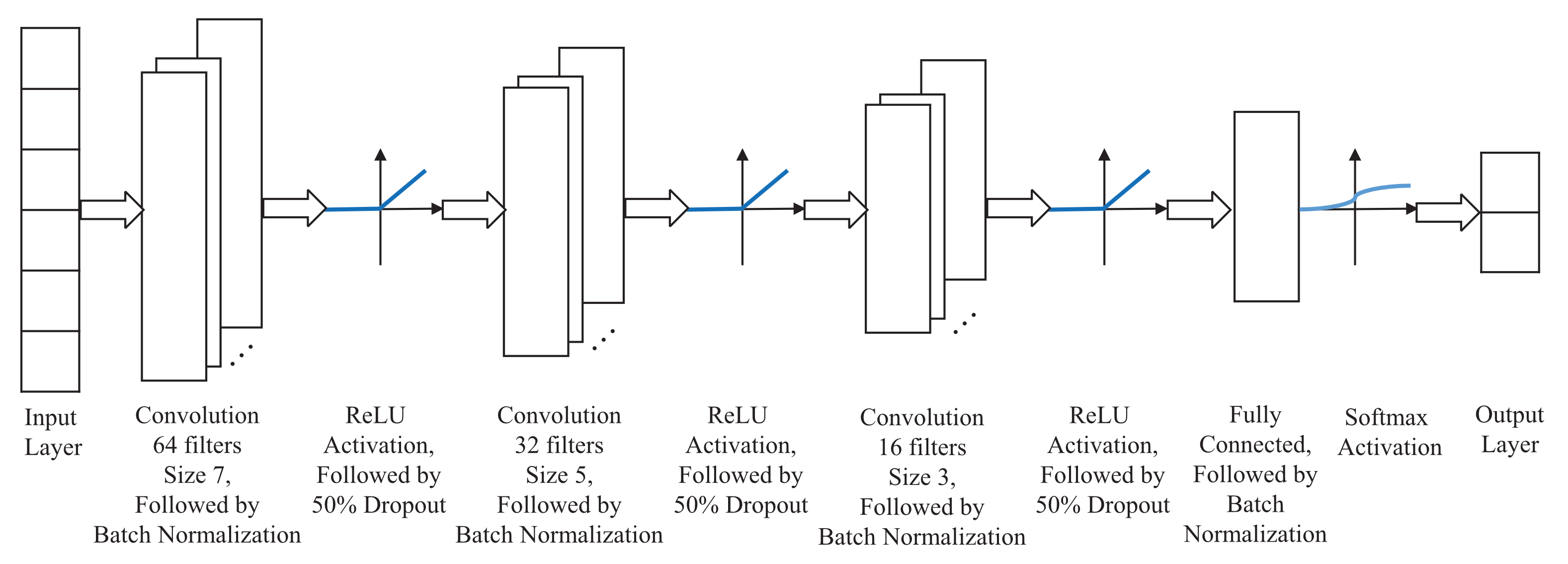}
    \caption{Architecture of the CNN used in \sysname}
    \label{fig:nn_zoom}
\end{figure*}

After the data preprocessing, we use a customized neural network to classify whether a potential gadget chain is benign or real. The architecture of our neural network (shown in Fig. \ref{fig:nn_zoom}) is a convolutional neural network (CNN), which is a feed-forward neural network that is good at learning data with spatial structures and capturing their local dependencies. Typically, a CNN has one input layer, multiple convolution layers, multiple activation layers, and one output layer. In particular, all of the convolution layer in the network is 1D convolution layer which is shown in Figure~\ref{fig:conv}.

The 1D convolution layer involves a mathematical operation called {\tt convolution}. 
Suppose $X$ represents the one-hot encoded instruction bytes, $w$ represents the convolution kernel (weight vector) whose length is $m$. Then the convolution between $X$ and $w$ is a matrix $X*w$, whose $ith$ column can be written as: 

\begin{equation}
    (X*w)_i = \sum_{j=1}^{m} X_{i-j+m/2} \cdot w_j
\end{equation}

The convolution aggregates information from adjacent bytes. This information includes certain byte values, the ordering of bytes, etc. The convolution layer is followed by a nonlinear activation layer (e.g., ReLU) to denote whether certain information is present or not. We stack three layers of convolution and activation to gradually capture higher level of information of the input bytes, e.g. the presence of a certain gadget, the ordering of different gadgets, the repetition of certain patterns, etc. The higher level of information is more abstract and difficult to extract or represent (similar to the case in image classification), but it is more related to the classification task, i.e., whether the chain is benign or real. The last activation layer is followed by a fully connected layer and another activation layer output to give a classification output, benign (0) or real (1). 

Since the input $X$ is fixed, only the weights $w$ influence the output. These values are not determined via heuristics or expert knowledge, as in many previous ROP detection methods. Instead, they are trained (recursively updated) by minimizing the differences between the true labels and the network's outputs. For details about CNN, please refer to \cite{goodfellow2016deep, kim2014convolutional}.

The architecture of our neural network is illustrated in Figure~\ref{fig:nn_zoom}, which is a zoom-in version of the CNN in Figure~\ref{fig_ropnn}. We first use a convolutional layer with 64 convolution filters with the kernel size of 7 (length of the convolution filter is 7). Then we perform batch normalization (BN) before a nonlinear activation function, which in this case is a rectified linear unit (ReLU). ReLU is a simple rectifier with the form of $f(x) = \max{(x, 0)}$. After the ReLU activation, we apply a 50\% dropout to prevent overfitting. Then we repeat this Convolution-BN-ReLU-Dropout structure for two more times. In particular, we use 32 convolutional kernels with the size of 5 and 16 convolutional kernels with size 3 for the next two layers. Then the output from the last Convolution-BN-ReLU-Dropout structure is feed into a fully connected layer. Finally, we use the softmax activation function as a classifier.\cut{ For the optimizer (the method to adjust all weights in the neural network), we use an optimization procedure called stochastic gradient descent (SGD) of learning rate 0.01 with momentum~\cite{qian1999momentum}.} It is worth mentioning that we do not include pooling layers which are widely used in image classification tasks, because \cut{for the reason that }our entire input is meaningful and downsampling our input vector would yield a completely different gadget chain. 

We use a grid search~\cite{hsu2003practical} to fine-tune the above configurable hyper-parameters (e.g., dropout rate, the filter sizes, etc) in our model and find the best set of values. Details are provided in Section~\ref{sec:evalution}. 


This network includes some modern techniques in deep learning such as BN \cite{ioffe2015batch} and dropout~\cite{srivastava2014dropout} to improve the performance. BN is a method to reduce internal covariance shift by normalizing the layer inputs in the neural network \cite{ioffe2015batch}, and has been shown to improve learning. Dropout is a simple way to prevent a neural network from overfitting by randomly dropping a percentage of neurons during the training phase. These methods have been widely used in recent years and seen great successes.

\subsection{Training CNN}\label{sec:train}
We train the CNN after collecting training samples (gadget-chain-like instruction sequences and real gadget chains) for a specific program. We use an optimization procedure called stochastic gradient descent (SGD) with learning rate 0.1 and momentum~\cite{qian1999momentum}. In particular, we input a mini-batch of the training samples into the network, then compute the outputs, the errors and the gradients for the input. The gradients are calculated by backpropagation to determine how the network should update its internal weights; e.g., $w_1$,$w_2$ and $w_3$ in Figure \ref{fig:conv}, which are used to compute the representation in each layer from the representation in the previous layer.\cut{ which computes gradients for algebraic expressions through recursive application of chain rule.} We repeat this process for many mini-batches until the errors converge or the maximum training epoch is reached. In practice, we also test another optimizer, i.e., Adam, and both the accuracy and the convergence speed are similar to those of the SGD, so we choose to use SGD as our optimizer. 

Note we set forth a goal (R2) to reach close to zero false positive rate for the \sysname. This is motivated by the fact that false positives can really irritate the system admin and undermine the usefulness of the detection method, or even impact its security. To reduce the false positive rate, we use a common technique, i.e., penalizing false positives more than false negatives. Specifically, we use the notation of ``penalizing factor'' to describe how much more a false positive contributes to the loss function compared to a false negative. A penalizing factor 1 means no difference between a false positive and a false negative, while a penalizing factor of 5 means a false positive contributes to the loss 5 times more than a false negative does. In the evaluation section, we empirically decide a value to achieve a very low false positive rate and a still high detection rate.

\section{Evaluation}\label{sec:evalution}


In this section, we evaluate \sysname on real-world programs. In particular, we want to answer the following questions. \textbf{1.} Does \sysname has high detection rates and low false positive rates for multiple programs? \textbf{2. }Can \sysname detect real-world ROP exploits? \textbf{3. }Can \sysname detect ROP exploits that can bypass other ROP detection methods, e.g., CFI? \textbf{4. }Is the speed of \sysname satisfiable? 

We train a CNN for each of the five tested programs, namely nginx 1.4.0, apache 2.4.18, proftpd 1.3.0a, vsftpd 3.03, ImageMagcik 7.08, respectively. The tested programs are widely adopted server-side applications (i.e., web servers, ftp servers, image processor). We compile, execute and memory-dump them on Ubuntu 16.04 32bit system. Note one specific CNN should be trained for one program because the (training + testing) dataset for one program is different from that for other programs. 
We collect and create 20 working ROP exploits for each of the five tested programs and \sysname successfully detects all of them. \sysname can also detect all 10 ROP exploits that bypass CFI. 




\subsection{How is Big Data Used to Generate Benign Training Dataset?}\label{sec: big_data}

Collecting abundant training data is one of the most challenging tasks when utilizing the neural network. We use the following three input datasets to generate the gadget-chain-like instruction sequences: (A) a medium size HTTP traffic dataset in~\cite{wang2010sigfree}; (B) all images in ImageNet~\cite{imagenet_cvpr09} (1.2 TB in total); (C) all PDF files from arXiv~\cite{arxiv} (800 GB in total). We use all (A)(B)(C) for the two web servers, i.e., nginx and apache, to do ASL-guided disassembly (described in Section~\ref{sec:asl}) and generate gadget-chain-like instruction sequences. Similarly, we use all (B) for the image processor (i.e., ImageMagick), and all (B)(C) for the two ftp servers (i.e., proftpd and vsftpd) to do the same job. As a summary, we consider all valid inputs for the target program. For example, all three datasets all typical traffic that pass web servers; image processor, however, only deals with images. 

To quickly process the huge amount of data, we use a Google Cloud Platform instance with 96 vCPUs. The memory dump size, input data, generation time, and the number of generated gadget-chain-like instruction sequences are shown in Table \ref{tab:generation}. As a brief summary, we generate 26k to 105k gadget-chain-like instruction sequences for different programs, respectively. 

It can be verified here that we do need a classifier in \sysname. Otherwise, if we simply do ASL-guided disassembly on input data and treat all potential gadget chains as real gadget chains, all gadget-chain-like instruction sequences generated above will become false positives. 



\begin{table*}[t]
  \centering
  \scriptsize
  \begin{tabular}{|c|M{12mm}|M{12mm}|M{10mm}|M{10mm}|M{10mm}|M{15mm}|M{30mm}|M{25mm}|c|}
  \hline
  Program & Memory Dump Size &  HTTP Data & ImageNet & arXiv PDF  &  Input Data Size & Generation Time & \# of Gadget-Chain-Like Instruction Sequences & \# of Real Gadget Chains \\ \hline
  nginx 1.4.0 & 3.2 MB & \checkmark & \checkmark & \checkmark & 2 TB & 7 hours & 40,674   & 40,674 \\ \hline
  apache 2.4.18 & 3.7 MB & \checkmark & \checkmark & \checkmark & 2 TB & 7 hours & 81,127  & 81,127 \\ \hline
  proftpd 1.3.0a & 2.4 MB & $\times$ & \checkmark & \checkmark & 2 TB & 7 hours & 26,020 & 26,020 \\ \hline
  vsftpd 3.03 & 4.5 MB & $\times$ & \checkmark & \checkmark & 2 TB & 7 hours & 105,057  &  105,057  \\ \hline
  ImageMagick 7.08 & 8.5 MB & $\times$ & \checkmark & $\times$ & 1.2 TB & 4 hours  & 46,224 & 46,224   \\ \hline
  \end{tabular}
 \caption{Generation of Gadget-Chain-Like Instruction Sequences}
\label{tab:generation}
\end{table*}

As explained in Section~\ref{sec:real}, for each program, we generate the same number of real gadget chains as gadget-chain-like instruction sequences. The generation is rapid and is done on a local workstation. Now that we have five datasets for the five programs, respectively. Each of them contains two smaller datasets: the gadget-chain-like instruction sequences and the real gadgets chains, which have the same number of samples.

\subsection{Tuning the hyper-parameters of \sysname}

Hyper-parameter tuning is a very important aspect of deep learning applications. Different hyper-parameter sets can lead to very different results. In this subsection, we use grid search~\cite{hsu2003practical} to empirically find the best hyper-parameters for our CNN. Grid search exhaustively searches the hyper-parameter space to look for the combination of hyper-parameters that yields the best result. In \sysname, we have a handful of hyper-parameters and the grid search works well. 

We implement our model using Keras~\cite{keras} and run it on NVIDIA Tesla P100 GPU. We ran the experiment under each setting for three times and take the average value of the accuracy. This averaged accuracy is the primary factor to compare different hyper-parameter settings. However, simply rely on accuracy leads to a problem. For example, let us assume 128 convolution filters already leads to the best result. Doubling the number and make it 256 is likely to be a waste of training time -- but the accuracy might see a tiny increase, say 0.05\%. So we also consider how fast the hyper-parameters lead to the convergence of error rate, measured in training epochs. If hyper-parameter set a is less than 0.05\% better than set b in accuracy, but at the same time the training epochs is 20\% more, we favor set b. In this way, we can not only get high accuracy, but also avoid unnecessary training time. 

We use the nginx dataset to fine-tune the hyper-parameters. We split the whole dataset into two subsets: a training dataset that is 80\% of the whole dataset (80\% of the samples from both labels), selected at random; a test dataset that is the rest 20\% of the samples. We use the training dataset to train the model and fine-tune the hyper-parameters.  

The parameters we aim to tune and their corresponding candidate values are listed in Table \ref{tab:param}. After a grid search, we find the best set of hyper-parameters, which is highlighted in Table~\ref{tab:param} (these are also the values included in Section~\ref{sec:nn}). We achieve 99.6\% accuracy under this best setting. 

\begin{table}[ht]
\centering
\scriptsize
\begin{tabular}{|l|l|}
\hline
Hyper-parameters & Candidate Values                            \\ \hline
\# of filters    & (32, 16, 8)\tablefootnote {This means the first layer has 32 filters, the second has 16, and the third has 8. The notation for filter size is interpreted in the same way}, (48, 24, 12), \textbf{(64, 32, 16)},    \\
                 & (96, 48, 24), (128, 64, 32)                 \\ \hline
filter size      & (3, 3, 3), (5, 3, 3), \textbf{(5, 5, 3)}, (7, 5, 3), \\
                 & (9, 5, 3), (9, 7, 5)                        \\ \hline
dropout rate     & 0.2, \textbf{0.5}, 0.8                      \\ \hline
learning rate    & 1, \textbf{0.1}, 0.01, 0.001                \\ \hline
\end{tabular}
\caption{Candidate values and the best value of hyper-parameters for \sysname}
\label{tab:param}
\vspace{-3mm}
\end{table}



We also consider using word embedding to preprocess the data, which works very well in NLP. But it does not help in our task. It produces similar accuracy to one-hot encoding but the required training time is 30\% longer. The longer training time can be explained by the fact the embedding layer itself needs to be trained along with the deep neural network. Also, instruction sequences are inherently different from language data, so the embedding is unable to improve the result. 

We do not use grid search to find the best penalizing factor (introduced in Section \ref{sec:train}). In fact, we not only want to minimize the false positive rate, we also want to maintain a high accuracy. However, using penalizing factor to decrease false positive rate inevitably increase false negative rate and may impact the overall accuracy, so a decision on the balance is better made manually. To illustrate this, we test an extremely large penalizing factor, 100, and the false positive rate is 0. However, the false negative rate is high and the overall accuracy drops to 92\%, which is undesirable. On the other hand, if we use a trivial penalizing factor 1, the false positive is 0.1\%. After manually testing several penalizing factors, we find a factor of 5 to be a good balance, where the false positive rate is 0.01\% and the detection rate is 99.3\%. Since we have same amount of samples for the two labels, so $accuracy =  (detection\;rate + (1 - false\;positive\;rate)) / 2$, thus the overall accuracy is still 99.6\%. 

A quick estimation demonstrates what 0.01\% false positive rate means in real world. Suppose a web server receives 10TB incoming traffic per day, since the number of potential gadget chains is rather small (40, 674 out of 2TB benign data), it is estimated only 20 false positives will be reported in 24 hours. Note the 10TB incoming traffic is equivalent to 1Gbps all the time during 24 hours, which is not a trivial amount, so the estimation is not under-estimating. 


\subsection{Can \sysname Accurately Detect Gadget Chains in the Test Dataset?}

In this subsection, we evaluate the accuracy of \sysname against the five commonly used programs.  

For every program, we select $80\%$ of the dataset as training data the rest 20\% as test data. We use the best hyper-parameters from the previous subsection, make 5-fold cross validation and the evaluation result is shown in Table \ref{tab:stat}. We reach 99.3\% detection rate and 0.01\% false positive rate for nginx 1.4.0, 98.7\% detection rate and 0.04\% false positive rate for apache 2.4.18, 98.3\% detection rate and 0.05\% false positive rate for proftpd 1.3.0a, 99.5\% detection rate and 0.02\% false positive rate for vsftpd 3.03, and 98.1\% detection rate and 0.02\% false positive rate for ImageMagick 7.08, respectively. As we can see from the results, \sysname has very high detection rates and very low false positive rates, and work well on different programs. 

The high accuracy shows that roughly 50k-100K samples are sufficient to train an accurate classifier for potential gadget chains. In contrast, for image classification, millions of samples are required to train neural networks~\cite{imagenet_cvpr09}. 

Besides CNN, there are also other candidate classifiers, e.g., RNN (LSTM), MLP and SVM. Our evaluation shows that the CNN is best suitable for this particular task. Due to space limitation, detailed comparisons are shown in Appendix \ref{app:comp}.

It is also noteworthy that the best hyper-parameters tuned on nginx dataset work well across different programs. This indicates although the specific instruction sequences are different in different programs (thus a dedicated CNN is needed), it has something in common across different programs. 

We observe that the training time roughly increases linearly proportional to the size of the training data. However, once the neural network is trained, the test speed is fast and less sensitive to the amount of test data. In the first row, it only takes 0.6 seconds to classify all test data for nginx (20\% of all samples, 16,270 in total). 

We also test what would happen if the amount of training data is small. In particular, we train the same model with 5\% of the nginx data (4K samples). We achieve 95.1\% detection rate and 1.2\% false positive rate. They are considerably inferior to the best achievable result. It validates that deep neural networks work better with larger training data since the networks can discover intricate features in large data sets instead of over-fitting small training data and missing key features. 

\begin{table*}[t]
  \centering
  \scriptsize
  \begin{tabular}{|c|c|M{15mm}|M{15mm}|M{15mm}|c|c|}
  \hline
  Program & CVE-ID & \# of Chains for Training & Detection Rate & False Positive Rate & Training Time & Classification Speed  \\ \hline
  nginx 1.4.0 & 2013-2028 & 65,078 & 99.3\% & 0.01\% & 45min  &  27,116 chains/s  \\ \hline
  apache 2.4.18 & N/A & 129,803 & 98.7\% & 0.04\% & 94min  &  27,120 chains/s  \\ \hline
  proftpd 1.3.0a & 2008-2950 & 41,632 & 98.3\% & 0.05\% & 31min  &  26,730 chains/s  \\ \hline
  vsftpd 3.03 & N/A & 168,091 & 99.5\% & 0.02\% & 118min & 26,020 chains/s \\ \hline
  ImageMagick 7.08 & N/A & 73,958 & 98.1\% & 0.02\%  & 55min & 27,231  chains/s  \\ \hline
  \end{tabular}
 \caption{Evaluation results on five tested programs}
\label{tab:stat}
\vspace{-3mm}
\end{table*}

\subsection{Can \sysname Accurately Detect Real-World ROP Exploits?}

In this subsection, we test \sysname against real-world ROP exploits that are collected in-the-wild or generated by ROP exploit generation tools (i.e., Ropper~\cite{ropper2016} and ROPC \cite{ropc}). 

The motivations behind these experiments are two-fold. Firstly, in the previous subsection, we show that the trained deep neural network of \sysname is an accurate classifier with a high detection rate and low false positive rate. However, despite our efforts to make the real gadget chains valid, they are not necessarily useful from an attacker's point of view. Therefore, we need a more direct evaluation here to show \sysname is able to detect real-world ROP exploits (that are not part of the training data). Secondly, since the real gadget chains are directly generated (not from ASL-guided disassembly), we also need to show the ASL-guided disassembly is capable of correctly identify the addresses of gadgets in a real gadget chain, which is the basis for our approach. 

Among the five tested programs, nginx and proftpd have known vulnerabilities and we directly exploit them with ROP attack. For the rest three, we use the latest version so there is no known vulnerability. Instead, we inject a trivial stack buffer overflow vulnerability into each of them to make the exploiting possible. In fact, as long as the exploit is a ROP attack, the underlying detail of the vulnerability has nothing to do with the effectiveness of our method, so the injected vulnerabilities do not undermine our evaluation. 

For each vulnerable program, we first obtain one ROP exploit that leads to a shell. For nginx, we use the attack script published in BROP \cite{HackingBlind}. For proftpd, we use the attack script published in Exploite-DB \cite{exploit-db} but change it to launch a ROP attack. For each of the rest three, we create a working exploit that leads to a shell. After that, we manually mutate the exploit to generate 4-5 more samples for testing. For example, we can exchange the order of several ROP gadgets without changing the behavior of the exploit. We also substitute gadgets with new gadgets that have the same effects. 

To further create test samples, we use Ropper \cite{ropper2016} to generate ROP exploits that execute mprotect or execve. Ropper can generate different exploits because we can block the used gadgets and force it to generate new ones. Meanwhile, we create ROP exploits with ROPC \cite{ropc}, which is a ROP compiler that can compile scripts written in a special type of language (ROP Language, ROPL) into a gadget chain. Although in Section \ref{sec:real} we are unable to use these tools to generate a large number of real gadget chains, it successfully generate sufficient amount of samples for us to test \sysname. 

To sum up, for each of the five vulnerable programs, we collect and create 20 ROP exploits. All of them are manually verified to be working. i.e., achieving their designed functionality, for example getting a shell or executing mprotect. 

We then test \sysname against all of the 100 exploits. We first observe that the ASL-guided disassembly successfully extracts the gadget addresses embedded in the payload and obtains the corresponding instruction sequences. Subsequently, the DNN correctly classifies all of the potential gadget chains as real gadget chains. This demonstrates the ASL-guided disassembler and the neural network synergize well and the system works as designed. \sysname is able to detect real-world ROP exploits. 

Thanks to the generalization capacity of DNN, the detection capacity of \sysname is not restricted to the ROP attacks it sees during training. In fact, it detects the blind ROP attack which is not part of the training data. Furthermore, ROP exploits generated by ROPC can have more than 100 gadgets, which are three times longer than the longest one in the training dataset. They are also correctly detected.

\subsection{Comparison between \sysname and CFI}

We compare \sysname with CFI, a real-world ROP detection (and prevention) method that is already being deployed in production systems. Specifically, we evaluate Bin-CFI, the CFI implementation proposed in \cite{zhang2013control}. 

Bin-CFI first disassembles the stripped binary and then statically analyze the direct and indirect call/jmp targets. Bin-CFI overcomes several difficulties and can eliminate 93\% of the gadgets in the program \cite{zhang2013control}. Bin-CFI is considered one of the strongest forms of CFI implementation \cite{davi2014stitching, goktas2014out}. 


As mentioned earlier, static point-to analysis is a very challenging task. Consequently, Bin-CFI cannot guarantee it always validates the control flow transfer targets precisely. Davi et al. \cite{davi2014stitching} observe that Bin-CFI does not (cannot) validate the integrity of the pointers in the global offset table (GOT) since they are initialized at runtime. They build a gadget chain to first maliciously overwrite a GOT entry and then invoke the desired function from the GOT, which is allowed by Bin-CFI. 

For each of the five program tested by us, we adopt the same trick to build two ROP exploits that can bypass Bin-CFI: one directly calls {\tt execve}, and another uses {\tt mprotect} to make the stack executable and then executes the shellcode from it. \sysname successfully detects all of the ten ROP exploits. 

The other pros and cons of \sysname and CFI are as follows. One advantage of \sysname over CFI is on the runtime overhead. CFI typically has a considerable runtime overhead, especially fine-grained CFI, since they have a stricter policy and require more checks. Conversely, \sysname is transparent to the protected program and does not incur any runtime overhead (please refer to the next subsection). Additionally, \sysname also does not need to instrument the program or modify the compiler. One advantage of CFI over \sysname is that CFI may stop the ROP exploit in action but \sysname is not a blocker. Nevertheless, IDS is designed to detect the attacks and enable other defense reactions. 


\subsection{Throughput of \sysname}

We now consider the throughput of \sysname on nginx. We use throughput, rather than latency, as the performance criteria because \sysname is an IDS and it does not block the traffic. In other words, \sysname exerts zero runtime overhead for the protected program. 

We can calculate from Table \ref{tab:generation} that the disassembler works at a speed of 665 Mb/s, which can match the traffic on a server that is not too busy. Moreover, since the ASL-guided disassembly can be  parallelized very well, we can split the workload across multiple servers running \sysname, further increasing the throughput. 



The CNN in \sysname can classify 27k potential gadget chains in one second. We observe the number of potential gadget chains is rather small. On average, in the 665 Mb data processed within one second, only 13 potential gadget chains are fed into the neural network for classification. Thus, the performance of the entire IDS is bounded by the speed of disassembly and the overall throughput is 665 Mb/s.

\section{Discussion and Limitations}\label{sec:discussion}

Readers may wonder how can \sysname work if the traffic is encrypted (e.g., HTTPS). Encryption can hide the addresses of ROP gadgets and hinder the ASL-guided disassembly. The solution is to deploy \sysname between the point of encryption and the protected program. For example, when a reverse proxy is deployed in front of a web-server to provide encryption, then \sysname should be deployed between the reverse proxy and the web-server. Now \sysname sees the unencrypted HTTP traffic and works in its normal way. 

Our IDS can cooperate well with Address Space Layout Randomization (ASLR). The two gadget chain datasets only contain instructions, not addresses. So the CNN is not affected by different memory layout. Meanwhile, when a server program is newly launched (thus a new memory space layout is created), we update the memory snapshot used by \sysname. In this way, the disassembler always works on the current image of the running program. So the result from the disassembler is also accurate. 
Given these two observations combined, we find that our IDS can work in conjunction with ASLR, further raising the bar for attackers. 

Note the memory space layout remains stable after the initial randomization (even in fine-grained ASLR). So the update does not happen very often and does not incur high runtime overhead (if any). Take nginx for an example, although several work processes are created, they are {\tt forked} from the master process and their memory layouts are the same. 

Readers may notice our criteria for a potential gadget in Section~\ref{sec:disassem} is relatively broad and it may allow non-gadgets. The reason is that we do not want to let any real ROP gadgets evade the disassembly engine and further bypass the detection. It makes our approach more robust and harder to bypass. But we showed this does not lead to too many false positives. 

In case of a patch or an update, whether the neural network needs to be re-trained depends on the detailed change of the program. If a part (e.g., a function) of the program is removed and this section happens to be present in some gadget chains, then these gadget chains should be removed from the dataset. If some instructions or functions are inserted into the program, it is very likely that the original gadget chains are still valid and there is no need to regenerate the dataset or re-train the neural network. 
Since training our neural network is not very time consuming, this kind of infrequent re-train is affordable.


ROP attack is getting more and more complex and has several variants. \sysname is able to detect polymorphic ROP attacks \cite{polyrop} because there always have to be some un-masqueraded ROP gadgets to unpack, decrypt, or arrange the real attacking gadgets. \sysname can detect these un-masqueraded ROP gadgets. On the other hand, some recent variations of ROP, e.g, JIT-ROP, that leverages a JavaScript environment to launch the attack, can potentially bypass our detection. However, unless the ROP attack is planned and launched ENTIRELY from JavaScript (which is considerably complex~\cite{JITROP}), we can still detect it. For example, if the attacker uses JavaScript to leak certain memory information or arrange the heap layout, and then launch ROP attack through the network, \sysname can still detect the ROP exploit. Meanwhile, high profile targets for ROP attacks, e.g., server programs, seldom provide a JavaScript environment.

\section{Conclusion}

\sysname is a novel intrusion detection system that leverages the power of deep neural networks to classify potential gadget chains produced by an ASL-guided disassembler. 
We show that \sysname has a high detection rate and a very low false positive rate. It also successfully detects all of the real-world ROP exploits collected or created by us. Meanwhile, it is non-intrusive and does not incur runtime overhead. We argue that \sysname is a practical widely-deployable detection method against ROP attacks. 

\newpage
\bibliographystyle{IEEEtran}
\bibliography{reference}

\begin{appendices} \label{app:comp}
\section{\sysname V.S. LSTM, MLP or SVM}

\begin{table*}[ht]
  \centering
  \scriptsize
  \begin{tabular}{|c|c|c|c|c|}
  \hline
  Method & Detection Rate & False Positive Rate & Training Time & Classification Speed  \\ \hline
  \sysname & 99.3\% & 0.01\%  & 45min & 27,116 chains/s \\ \hline
  ASL-guided \& LSTM & 97.8\% & 0.2\% & 143min  &  7,150 chains/s \\ \hline
  ASL-guided \& MLP & 96.9\% & 22.2\% & 26min  &  85,629 chains/s \\ \hline
  ASL-guided \& SVM & 77.6\% & 42.4\% & 12min & 310 chains/s \\ \hline

  \end{tabular}
 \caption{Comparison of \sysname, LSTM, MLP and SVM on nginx}
\label{tab:comparison}
\end{table*}

In this appendix, we compare the performance of \sysname to that of combining ASL-guided disassembly with other well-known classifiers, i.e., LSTM (one type of RNN), Multiple-Layer Perceptron (MLP) and Support Vector Machine (SVM). These three competitors are also widely used and produce many good results~\cite{mountrakis2011support, summers2016progress, yao2015describing}.  Specifically, we use an LSTM with 96 hidden units, an MLP with three fully-connected layers with 32 units in each layer, and an SVM with the radial basis function (RBF) kernels, and  And we use 80\% of Nginx dataset for training. The results are shown in Table \ref{tab:comparison}.

The combination of ASL-guided disassembly and MLP leads to 96.9\% detection rate and 22.2\% false positive rate. The combination of ASL-guided disassembly and SVM leads to 77.6\% detection rate and 42.4\% false positive rate. These two false positive rates are too high for practical application because they can easily cause denial of service. LSTM reaches a 97.8\% detection rate and 0.2\% false positive rate, which is still far from CNN's result. Also, LSTM takes 3 times longer to train. 

The result is not surprising. In fact, RNN (or LSTM) is more suitable for temporal data \cite{goodfellow2016deep} whereas our instruction sequence data is more spatial in nature. MLP uses fully connected layer that is known to be hard to train. Using the same amount of data, it does not achieve comparable accuracy to CNN. The SVM is not good at directly dealing with minimally preprocessed data and it requires certain feature engineering to first extract features from the input data. In summary, CNN is superior to its three competitors and more suitable to the task. 

\end{appendices}

\end{document}